\documentclass[useAMS,usenatbib]{mn2e}

\newcommand{\be}{\begin{equation}}
\newcommand{\ee}{\end{equation}}
\newcommand{\ba}{\begin{eqnarray}}
\newcommand{\ea}{\end{eqnarray}}
\newcommand{\asec}{\prime\prime}

\include{epsf}

\title[The Cosmic Horseshoe Lens]
{Models of the Cosmic Horseshoe Gravitational Lens J1004+4112}
\author[S. Dye et al.]
{S. Dye$^{1}$\thanks{E-mail:s.dye@astro.cf.ac.uk},
N. W. Evans$^{2}$,
V. Belokurov$^{2}$,
S. J. Warren$^{3}$,
P. Hewett$^{2}$\\
$^{1}$School of Physics and Astronomy, Cardiff University,
Queens Buildings, The Parade, Cardiff, CF24 3AA, UK\\
$^{2}$Institute of Astronomy, University of Cambridge, Madingley Road,
Cambridge CB3 0HA, UK \\
$^{3}$Astrophysics Group, Imperial College London, Blackett Laboratory,
Prince Consort Road, London, SW7 2AZ, UK \\
}

\begin{document}

\date{}

\pagerange{\pageref{firstpage}--\pageref{lastpage}} 
\pubyear{2008}

\maketitle

\label{firstpage}

\begin{abstract}
We model the extremely massive and luminous lens galaxy in the Cosmic
Horseshoe Einstein ring system J1004+4112, recently discovered in the
Sloan Digital Sky Survey. We use the semi-linear method of Warren \&
Dye (2003), which pixelises the source surface brightness
distribution, to invert the Einstein ring for sets of parameterised
lens models. Here, the method is refined by exploiting Bayesian
inference to optimise adaptive pixelisation of the source plane and to
choose between three differently parameterised models: a singular
isothermal ellipsoid, a power law model and a NFW profile. The most
probable lens model is the power law with a volume mass density
$\rho\propto r^{-1.96\pm0.02}$ and an axis ratio of $\sim 0.8$. The
mass within the Einstein ring (i.e., within a cylinder with projected
distance of $\sim 30$ kpc from the centre of the lens galaxy) is
$(5.02\pm0.09)\times10^{12}\, {\rm M}_\odot$, and the mass-to-light
ratio is $\sim 30$.  Even though the lens lies in a group of galaxies,
the preferred value of the external shear is almost zero. This makes
the Cosmic Horseshoe unique amongst large separation lenses, as almost
all the deflection comes from a single, very massive galaxy with
little boost from the environment.
\end{abstract}


\section{Introduction}

The measurement of galaxy mass distributions using strong
gravitational lensing is now a well-established process, having found
application to several tens of systems to date \citep[for example,
see] [and references therein]{dye07}. The main attraction of strong
lensing over other methods is its insensitivity to the dynamical state
of the deflecting mass.The main disadvantage is that some
features of the lens mass distribution, such as the ellipticity, are
much more robustly constrained by the modelling than others, such as
the radial profile \citep{Sa03}.

Multiple images of a background source can constrain the radial
profile of the lens projected mass density only weakly \citep[for
example, see the review by][]{schneider06}.  However, some of the
degeneracy is lifted by the incorporation of extra constraints from
the observed velocity dispersion profile of the lens, a technique
first applied by \citet{sand02} to the cluster MS 2137$-$23 and by
\citet{treu02} to the early type galaxy MG 2016+112 and subsequently
to a number of systems since \citep{koopmans03,sand04}.

\citet{dye05} showed how Einstein ring systems, i.e., strong lens
systems where an extended source is imaged into a complete or
near-complete ring, can constrain the mass profile of the lens more
strongly than systems with multiple point-like images. This work used
the semi-linear method of \citet{warren03}, so called because the
problem of finding the best fit lens model and source surface
brightness distribution is split into a linear inversion of the source
for a given non-linearly parameterised lens model. The technique has
been used by several other studies to date
\citep{treu04,treu06,koopmans06}.  \citet{koopmans05} presented an
enhanced version of the method which also reconstructs the lens
gravitational potential non-parametrically. In addition, a Bayesian
version of the semi-linear method was developed by \citet{suyu06}.

In this paper, we apply the semi-linear method to reconstruct the lens
mass profile and source surface brightness image of the Cosmic
Horseshoe Einstein ring system J1004+4112, recently discovered in the
Sloan Digital Sky Survey by \citet{belokurov07}. This is one of the
largest and most complete Einstein rings thus far discovered, with a
diameter $10^{\asec}$ and subtending an angle of $\sim 300^\circ$.
The lens is an exceptionally massive Luminous Red Galaxy (LRG) with a
redshift of 0.44 and a velocity dispersion of $\sim 430$ kms$^{-1}$,
estimated from a mediocre signal-to-noise spectrum.  The source is a
star-forming galaxy of BX type, using the nomenclature of
\citet{St04}, with a redshift of 2.379.

\citet{belokurov07} already provided some simple analysis, by picking
out four density knots or maxima in the ring and using techniques from
the modelling of quadruply-imaged point sources to reconstruct the
lensing mass \citep{Ev03}.  This modelling threw up a number of
unresolved questions. First, there are more than four density maxima
in the ring, hence \citet{belokurov07} provided a number of
possibilities for the mass reconstruction. Their models were
restricted to scale-free, isothermal-like mass profiles, though with
rather general azimuthal variations. The origin of the additional
density maxima in the ring was unclear -- they were thought to arise
from the lensing of more than one source or from higher order
(sextuple) imaging. Second, although the LRG lies in a galaxy group,
the group's contribution to the lensing deflection via external shear
was found to be modest. Apparently, almost all of the lensing effect
is provided by the LRG itself. This is surprising because almost all
the known lenses with image separations greater than $\sim 3^{\asec}$
are produced by over-dense environments, with a significant lensing
enhancement provided by the group or cluster. Third, although the
visible light distribution of the LRG is nearly circular, the mass
reconstructions were more flattened and irregular.  Fourth, although
\citet{belokurov07} provided models that matched the image location,
they did not successfully reproduce the image magnifications.  All
this motivates a return to the Cosmic Horseshoe, but with a more
sophisticated ring modelling technique.

Here, we determine the most probable mass profile for the
Cosmic Horseshoe lens from three popular models. This is done by using
a refinement of the semi-linear method of \citet{warren03}.  To compare
between models, we follow the technique of maximising the Bayesian
evidence as derived by \citet{suyu06}.  The layout of this paper is
as follows. In Section \ref{sec_data}, we briefly describe the
data. Our method of analysis is outlined in Section
\ref{sec_sl_method} and applied in Section \ref{sec_results}.  We
summarise the findings of this work in Section \ref{sec_summary}.
Throughout this paper, we assume the following cosmological
parameters; ${\rm H}_0=70\,{\rm km\,s}^{-1}\,{\rm Mpc}^{-1}$,
$\Omega_m=0.3$, $\Omega_{\Lambda}=0.7$.

\section{Data}
\label{sec_data}

The Cosmic Horseshoe was discovered by \citet{belokurov07} by
searching the Sloan Digital Sky Survey (SDSS) for luminous red
galaxies with multiple, faint, blue companions. The centre of the lens
galaxy lies at ($11^h48^m33.15^s,19^{\circ}30'3.5''$).  We refer the
reader to this discovery paper for full details of the data and
reduction which we briefly outline here.

Follow-up imaging of the lens system was carried out in May 2007 at
the 2.5m Isaac Newton Telescope (INT) in La Palma. Images were
acquired in the wavebands $U$, $g$ and $i$ with the Wide Field
Camera. Each image was integrated for a total of 600s and reduced with
the Cambridge Astronomical Survey Unit INT pipeline
\citep{irwin01}. The data in each band are shown in the first row of
Figure \ref{recon}.

Long-slit spectroscopy of the lens galaxy and arc was also carried out
in May 2007 at the 6m BTA telescope of the Special Astrophysical
Observatory (SAO), Nizhnij Arkhyz, Russia.  Absorption by Ca, H and K
in the lens spectrum places the lens galaxy at a redshift of $z=0.44$.
\citet{belokurov07} estimate a velocity dispersion of the lens of
$430\pm50$ kms$^{-1}$ by Gaussian profile fitting to the absorption
lines. The slit was placed $\sim 1''$ from the centre of the lens
which, given the seeing of $1.7''$ and effective radius of $\sim 2''$,
means that the spectrum is dominated by flux from within the half
light radius.  Ly$\alpha$ emission and absorption features in the
spectrum of the arc indicate that the source lies at a redshift of
$z=2.38$.

To remove possible contamination of the ring by flux from the lens
galaxy, we fitted an elliptical Sersic profile to the lens galaxy in
each waveband.  The fitted profiles were subtracted prior to our
analysis. The second row in Figure \ref{recon} shows the lens removed
ring image for each waveband.

Table \ref{tab_lens_light} lists the $U$, $g$ and $i$ best
fit parameters of the Sersic profile which has the form
\be
{\rm L}={\rm L}_{1/2}\exp\{-B(n)[(r/r_0)^{1/n}-1]\} \, .
\ee 
The parameters ${\rm L}_{1/2}$, $n$ and $r_0$ were allowed to vary in
the fit as well as the axis ratio, $q_s$, (i.e., minor axis divided by
major axis), orientation, $\theta_s$, and the centroid.  We use the
expression for $B(n)$ given by \citet{ciotti99}. In the fitting, we
convolved each trial surface brightness profile with a Gaussian point
spread function (PSF) that matched the image seeing determined from
stars in the field.  All three fits gave acceptable $\chi^2$
values. Note that the ellipticity and position angle of the major axis
are in good agreement with the results in Table~1 of
\citet{belokurov07}, who fitted a PSF-convolved de Vaucouleurs profile
to the light distribution.

\begin{table}
\centering
\small
\begin{tabular}{cccc}
\hline
Parameter & $U$ & $g$ & $i$ \\
\hline
L$_{1/2}$              & $1.20\pm0.32$ & $6.9\pm0.3$   & $61.2\pm0.4$  \\
$n$                    & $4.24\pm0.4$  & $4.71\pm0.12$ & $5.40\pm0.04$ \\
$r_0$ $(\asec)$ & $1.5\pm2.0$   & $6.1\pm0.6$   & $3.9\pm0.1$   \\
$\theta_s$ $(^\circ)$  & $85\pm20$     & $91\pm4$      & $91\pm1$      \\
$q_s$                  & $0.92\pm0.12$ & $0.83\pm0.03$ & $0.88\pm0.01$ \\
\hline
\end{tabular}
\normalsize
\caption{Sersic profile parameters fit to the $U$, $g$ and $i$ band
data. The normalisation L$_{1/2}$ has units of image counts 
matching the observed images in Figure \ref{recon}. The orientation
$\theta_s$ is in degrees counter-clockwise with respect to the
positive y-axis.}
\label{tab_lens_light}
\end{table}

\section{Methodology}

\label{sec_sl_method}

\subsection{Bayesian Semi-linear Inversion}

The original semi-linear method was derived by \citet{warren03}, first
applied by \citet{dye05} and placed within a Bayesian framework by
\citet{suyu06}. We give an outline of the method in this section but
refer the reader to these publications for more comprehensive details.

The technique assumes a pixelised image and source plane.  The term
`semi-linear' refers to the fact that the inversion problem can be
divided into a set of linear parameters -- the surface brightnesses of
the source plane pixels -- and a set of non-linear parameters that
define the lens model. Generally, the source surface brightness
distribution must be regularised to ensure that the linear inversion
step is not mathematically ill-posed (see below). This gives rise to
an extra non-linear parameter called the {\em regularisation weight}. 

\citet{warren03} noted that heavy regularisation biases the
reconstructed source, in turn biasing the best fit lens
model. Therefore, instead of applying regularisation, \citet{dye05}
ensured a well posed linear inversion through use of an adaptively
gridded source plane. In this way, regions of the source plane that
are not well constrained by the observed ring, i.e., areas of low
magnification, are gridded with large pixels whilst strongly
constrained areas of the source plane are more finely gridded.  The
degree to which source pixel sizes depend on the magnification is
controlled through another non-linear parameter called the {\em
splitting factor} (see below). In addition to ensuring a well posed
problem, an adaptive grid has the appealing characteristic that the
reconstructed source has a more uniform error map.

A more serious problem with regularisation is that it smoothes the
reconstructed source, effectively increasing the number of degrees of
freedom by an amount that cannot be satisfactorily quantified. This is
especially problematic when comparing different lens models, as a
fixed regularisation weight for one model generally does not give the
same increase in number of degrees of freedom for another. Therefore,
when comparing different regularised models, $\chi^2$ is
not a useful statistic.

In the Bayesian version of the semi-linear method derived by
\citet{suyu06}, the regularisation weight is set automatically by the
data. Crucially, the problem of comparing different lens models is
solved by the Bayesian evidence which allows models to be objectively
ranked as we describe below.

In the present work, we combine the advantages of both the Bayesian
approach {\em and} an adaptive source grid. As well as allowing model
ranking and regularisation, the Bayesian evidence lets the data
select the optimal source pixelisation by finding the most
probable splitting factor.

In the analysis outlined in the next section, it is helpful to keep
the regularisation weight and splitting factor segregated from the
linear source surface brightnesses and the non-linear lens model
parameters. Following the terminology of \citet{barnabe07}, we will
refer to these extra two non-linear parameters as 'hyperparameters' by
virtue of their indirect influence on the lens and source.

\subsection{Implementation of the Inversion Method}
\label{sec_imp}

The process of establishing the most probable lens parameterisation is
split into three levels of inference. In the innermost level, the best
fit source surface brightness distribution for a given set of lens
model parameters and hyperparameters is determined with a linear
inversion step. This proceeds as follows: A PSF-smeared image is
computed for every source pixel. All images are created using unit
surface brightness source pixels. The linear problem of finding the
factor required to scale each image such that their co-addition best
fits the observed image gives the best fit source pixel
surface brightnesses, which as a vector is \citep{warren03}
\be
\label{eq_sl}
\mathbf{s} = ({\rm \mathbf{F}} +
\lambda {\rm \mathbf{H}})^{-1} \mathbf{c} \, .
\ee
The square matrix ${\rm \mathbf{F}}$ and the vector
${\rm \mathbf{c}}$ have the elements
\be
{\rm F}_{ik}=\sum_j f_{ij}f_{kj}/\sigma_j^2 \, , \,\,\,
c_i=\sum_j f_{ij}d_j/\sigma_j^2
\ee
and $\mathbf{s}$ is a vector containing the best fit source pixel
surface brightnesses.  Here, $d_j$ is the observed flux in image pixel
$j$, $\sigma_j$ its error and $f_{ij}$ is the flux in pixel $j$ of the
image of source pixel $i$ for the current lens model.  The solution is
regularised by the square regularisation matrix ${\rm \mathbf{H}}$,
scaled by the regularisation weight $\lambda$ \citep[see][and Warren
\& Dye 2003]{press01}.  The standard errors of the reconstructed
source pixels are given by the diagonal terms of the covariance matrix
$\mathbf{C}$ which is just
\be
\label{eq_cov_matrix}
\mathbf{C}=({\rm \mathbf{F}} + \lambda {\rm \mathbf{H}})^{-1}\, .
\ee

In Bayesian terminology, computing the solution for $\mathbf{s}$ using
equation (\ref{eq_sl}) amounts to finding the most likely source
surface brightness distribution by maximising the posterior
probability for a given lens model and a given source pixelisation and
regularisation.

In the second level of inference, the most probable set of
hyperparameters for a given lens model is determined by maximising the
Bayesian evidence. The evidence is a probability distribution in the
lens parameters and hyperparameters that normalises the Bayesian
expression for the posterior probability.  It allows different models
to be ranked to find the most probable model (see below). 
\citet{suyu06} derived the evidence, $\epsilon$, for this
problem, which in our case can be expressed as
\ba
\label{eq_evidence}
-2 \,{\rm ln} \, \epsilon &=& 
\sum_j \left[\frac{\sum_i s_i f_{ij}-d_j}{\sigma_j}\right]^2
+{\rm ln} \, \left[ {\rm det} (\mathbf{F}+\lambda\mathbf{H})\right]
\nonumber \\
& & -{\rm ln} \, \left[ {\rm det} (\lambda\mathbf{H})\right]
+ \lambda\mathbf{s}^{T}\mathbf{H\,s} +\sum_j {\rm ln} (2\pi \sigma_j^2) \, 
\ea
where the summations in $j$ act over all image pixels and the
summation in $i$ acts over all source pixels. Here, we have assumed
zero covariance between all image pixel pairs.  In this expression,
the first term corresponds to $\chi^2$ and the fourth term regularises
the solution \citep[the term denoted $\lambda G_L$ in the work
of][]{warren03}.  In this second level, equation (\ref{eq_sl}) must be
evaluated for every trial set of hyperparameters to allow 
calculation of the evidence via equation (\ref{eq_evidence}).

Finally, in the third and outermost level of inference, the most
probable lens parameters are determined by maximising the evidence
obtained from the second level. Formally, to rank models, the evidence
should first be marginalised over the hyperparameters.  However,
\citet{suyu06} noted that a reasonable simplification is to
approximate the distribution function of the hyperparameters as a
delta function so that the maximum of the evidence obtained in the
second level can be directly compared between models.  We have adopted
this approximation in the present study.

In practical terms, the three-level procedure can be simplified. As
\citet{barnabe07} point out, the hyperparameters that maximise the
evidence in the second level of inference vary only slightly with
different trial lens model parameters in the third level. This means
that it is not necessary to maximise the hyperparameters with every
trial lens parameter set. Instead, we alternate between varying the
lens parameters whilst keeping the hyperparameters fixed and varying
the hyperparameters whilst keeping the lens parameters fixed. We start
this process by holding the hyperparameters (i.e., the regularisation
weight and splitting factor) at a large value and varying the lens
model. This reduces local maxima resulting in a smoother evidence
surface so that an initial set of lens parameters lying close to the
global maximum can be efficiently found \citep[see also][]{warren03}.

We note two further practicalities. First, when computing $\chi^2$, i.e.,
the first term in equation (\ref{eq_evidence}), we carry out the sum
over pixels contained within an annular mask that surrounds the ring.
The mask is designed to include the image of the entire source plane,
with minimal extraneous sky. This means that only significant image
pixels are used, making the evidence more sensitive to the model
parameters. Second, we use a simulated annealing downhill simplex
minimisation algorithm to minimise ${\rm -ln}\,\epsilon$ given by
equation (\ref{eq_evidence}). We find that a slow exponentially cooled
temperature with a half-life of $\sim 30$ iterations works extremely
well in finding the desired minimum.

\subsection{Adaptive Source Plane Grid}
\label{sec_adaptive_gridding}

We adaptively grid the source plane according to the prescription
given in \citet{dye05} and \citet{dye07b}. In this scheme, smaller
pixels are concentrated in higher magnification regions where there
are stronger constraints per unit area of the source plane.

The adaptive gridding algorithm starts with a regular
mesh of large pixels. The average magnification $\mu_i$ of every
source pixel $i$ is then computed. Those pixels that meet the criterion
$\mu_i \, r_i \geq s$ are then split into quarters, where $r_i$ is the
ratio of the area of pixel $i$ to the area of an image pixel and $s$
is the `splitting factor'. Having finished the initial loop through
all pixels, the process is repeated for the sub-pixels, then for the
sub-sub-pixels and so on until all pixels satisfy the splitting
criterion. 

The procedure is carried out every time the splitting factor is varied
in the evidence maximisation. Although the adaptive grid is dependent
on the lens model, we find that it does not vary significantly when
varying the lens parameters. Therefore, as we discussed in the
previous section, similar to the regularisation weight, we hold the
splitting factor fixed whilst varying the lens parameters and vice
versa, alternating until convergence is achieved. Convergence
typically occurs after only a few alternate loops.

\citet{suyu06} advocate second order regularisation, whereby source
surface brightness distributions that exhibit strongly varying
gradients are penalised more heavily than those with more gradual
gradient changes. Although this is simple to implement with a regular
source pixel grid, it is ill-defined on an adaptive grid (one
can't define a set of co-linear pixel triplets). Instead, we
apply a form of first order regularisation where strongly varying
pixel-to-pixel surface brightnesses are penalised.  We construct a
matrix that takes the difference between a given pixel $k$ and the sum
of all neighbouring pixels $l$ weighted by ${\rm
w}_{kl}=(a_l/a_k)\,N\exp(-y_{kl}^2/2\sigma^2)$. Here, $a$ is the
source pixel area, $y_{kl}$ is the separation of the centres of pixels
$k$ and $l$ and $N$ is a normalisation constant set such that
$\sum_{l,l\ne k} {\rm w}_{kl}=1$ and $N=1$ when $k=l$. We set
$\sigma=0.2''$. The matrix $\mathbf{w}$ here relates to the
regularisation matrix ${\rm \mathbf{H}}$ via ${\rm
\mathbf{H}}=\mathbf{w}^{T}\mathbf{w}$. By construction, $\mathbf{H}$
is non-singular so that the third term in equation (\ref{eq_evidence})
is always calculable.

In Section \ref{sec_results}, we show how each lens model prefers a
different value of the splitting factor, and how this couples to the
regularisation weight.

\subsection{Lens models}

We consider three popular mass profiles to model the distribution of
the total (baryonic and dark) projected lens mass: 

\begin{itemize}

\item {\em Singular isothermal ellipsoid (SIE)} -- This model has been
widely used in gravitational lensing \citep[see e.g.,][]{Ka93,Sc92}
motivated by a wealth of stellar dynamical evidence favouring the idea
that galaxies are nearly isothermal.  The projected surface mass
density follows $\kappa = \kappa_0 ({\tilde r}/1{\rm kpc})^{-1}$,
where ${\tilde r}$ is the elliptical radius defined by ${\tilde
r}^2=x^{\prime2}+q^2y^{\prime2}$.  The coordinates $x^\prime$ and
$y^\prime$ are defined on axes aligned with the semi--major and
semi--minor axes of the ellipse and $q$ is the ratio of the
minor to the major axis. There are a total of five parameters
for the SIE model: the normalisation $\kappa_0$, the orientation
$\theta$, the axis ratio $q$ , and the lens centroid in the image
plane, $(x_c,y_c)$.

\item {\em Navarro, Frenk \& White (NFW) profile} -- This profile was
introduced by \citet{nfw96} as a fit to dark matter halos created in
cosmological N-body simulations. The lensing properties have been
discussed by a number of authors \citep[see e.g.,][]{Ba96, Ev98,
keeton02}.  It has a projected surface mass density given by
\be
\kappa=\kappa_0\frac{1-F(x)}{x^2-1} 
\ee 
where $x={\tilde r}/r_s$ and 
\be F(x)
= \cases{ {1 \over \sqrt{x^2-1}}\,\mbox{tan}^{-1} \sqrt{ x^2-1 } &
$(x>1)$ \cr {1 \over \sqrt{1-x^2}}\,\mbox{tanh}^{-1}\sqrt{ 1-x^2 } &
$(x<1)$ \cr 1 & $(x=1)$ \cr } 
\ee 
The model is described by six parameters, but we vary five in the
evidence maximisation, keeping the scale radius $r_s$ fixed at the
value of 110kpc ($\equiv 20^{\asec}$ at $z=0.44$). This is in
accordance with the prediction by \citet{bullock01} for a galaxy of
similar mass and redshift to the cosmic horseshoe lens.  As has been
shown elsewhere \citep{dye07b}, the lensing properties of the NFW
profile depend only weakly on the value of $r_s$ assumed, with a 10\%
change in $r_s$ giving rise to only a $\sim 1\%$ change in the best
fit model parameters. The five parameters varied in the maximisation
are therefore: lens normalisation $\kappa_0$, orientation $\theta$,
axis ratio $q$, and lens centroid in the image plane $(x_c,y_c)$.

\item {\em Power law (PL)} -- This family of models was introduced by
\citet{Ka93}. The projected surface mass density is stratified on
concentric ellipses following the power-law form
$\kappa=\kappa_0\,({\tilde r}/{\rm 1kpc})^{1-\alpha}$.  The SIE is the
special case $\alpha=2$. The model has six parameters: lens
normalisation $\kappa_0$, orientation $\theta$, axis ratio $q$,
power-law slope $\alpha$ and lens centroid in the image plane
$(x_c,y_c)$.

\end{itemize}

For each model, we maximise the evidence with and without an external
shear component. The external shear adds a further two parameters to
each model, a magnitude $\gamma$ and an orientation
$\theta_\gamma$. The deflection angle required in the ray tracing has
an analytic form for the SIE model, but must be numerically evaluated
for the NFW and PL models, using the prescription given by
\citet{keeton02}.

We note at this point a common misconception regarding the mass sheet
degeneracy \citep[e.g.,][]{gorenstein88}. The degeneracy is such that
image structures are invariant under the transformation $\kappa
\rightarrow 1-a \, + \, a\,\kappa$ where $a$ is a constant. The
degeneracy is only applicable to lens models that remain self-similar
under the transformation. None of the three models applied in this
paper falls into this category. For instance, applying the
transformation to the power-law model does not produce a new
power-law. Inverting the argument, this means that no combination of
power-law parameters can give a model with a homogeneous sheet of
matter and in this sense, the mass sheet degeneracy is eliminated.

\section{Results}
\label{sec_results}

Table \ref{tab_params} shows the maximised parameters for the three
lens models with and without external shear using the $g$ band
data. The most probable model is the power law with a slope of
$1.96\pm0.02$. The evidence ranks the SIE as the next most probable
model, being only 10\% as probable as the power law.  Finally, the NFW
is strongly rejected, being ranked $>10^{-10}$ times less probable
than the power law. This is perhaps not surprising given that the
NFW is derived from simulations that neglect the effect of baryons.

\begin{figure*}
\epsfxsize=17cm
{\hfill
\epsfbox{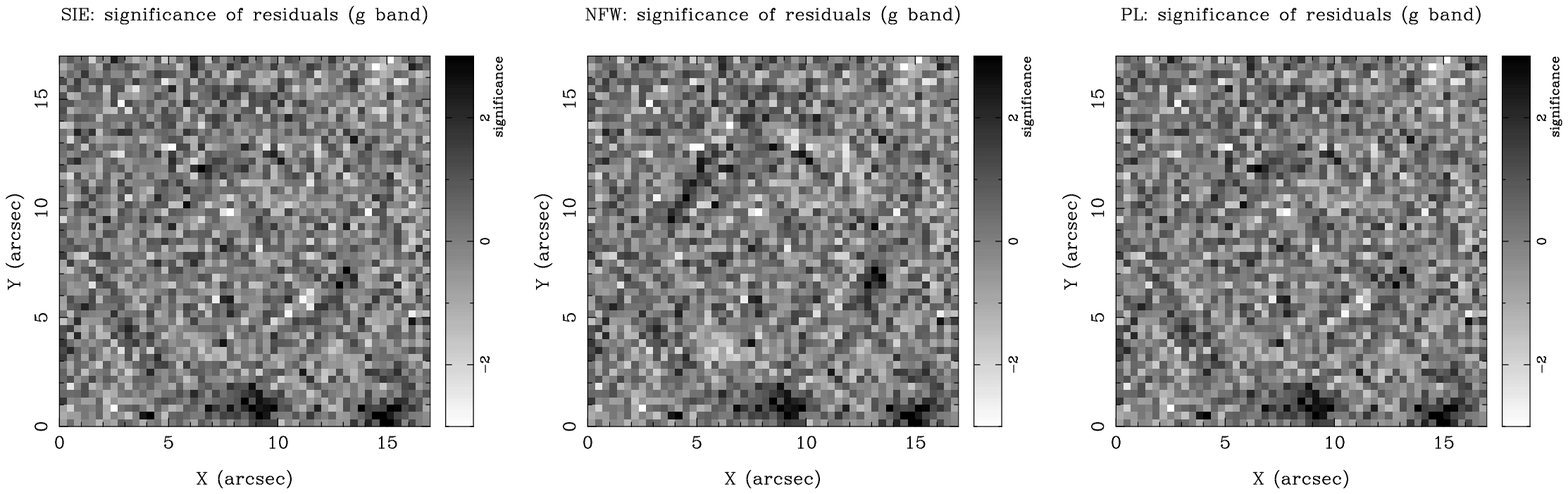}
\hfill}
\epsfverbosetrue
\caption{Significance of the residuals left after subtracting the
lensed image of the reconstructed source from the observed $g$ band
ring for the SIE, PL and NFW.}
\label{resids_cf}
\end{figure*}

Figure \ref{resids_cf} shows the significance of the residuals that
remain after subtracting the lensed image of the reconstructed source
from the observed $g$ band ring for the SIE, PL and NFW. The
NFW clearly leaves the strongest residuals as one would expect
from the evidence. The difference between the PL and SIE residuals
is not obvious upon visual inspection, however they differ with an
RMS of $\sim 5\%$.

For the best-fitting PL model, the mass within the
Einstein ring (i.e., within a cylinder with projected distance of
$\sim 30$ kpc from the centre of the lens galaxy) is
$(5.02\pm0.09)\times10^{12}\, {\rm M}_\odot$, as much as the entire
Local Group. Using the absolute magnitude in the $r$ band of $-23.45$
computed by \citet{belokurov07}, the mass-to-light ratio is $\sim 30$.

\begin{figure*}
\epsfxsize=11.1cm
{\hfill
\epsfbox{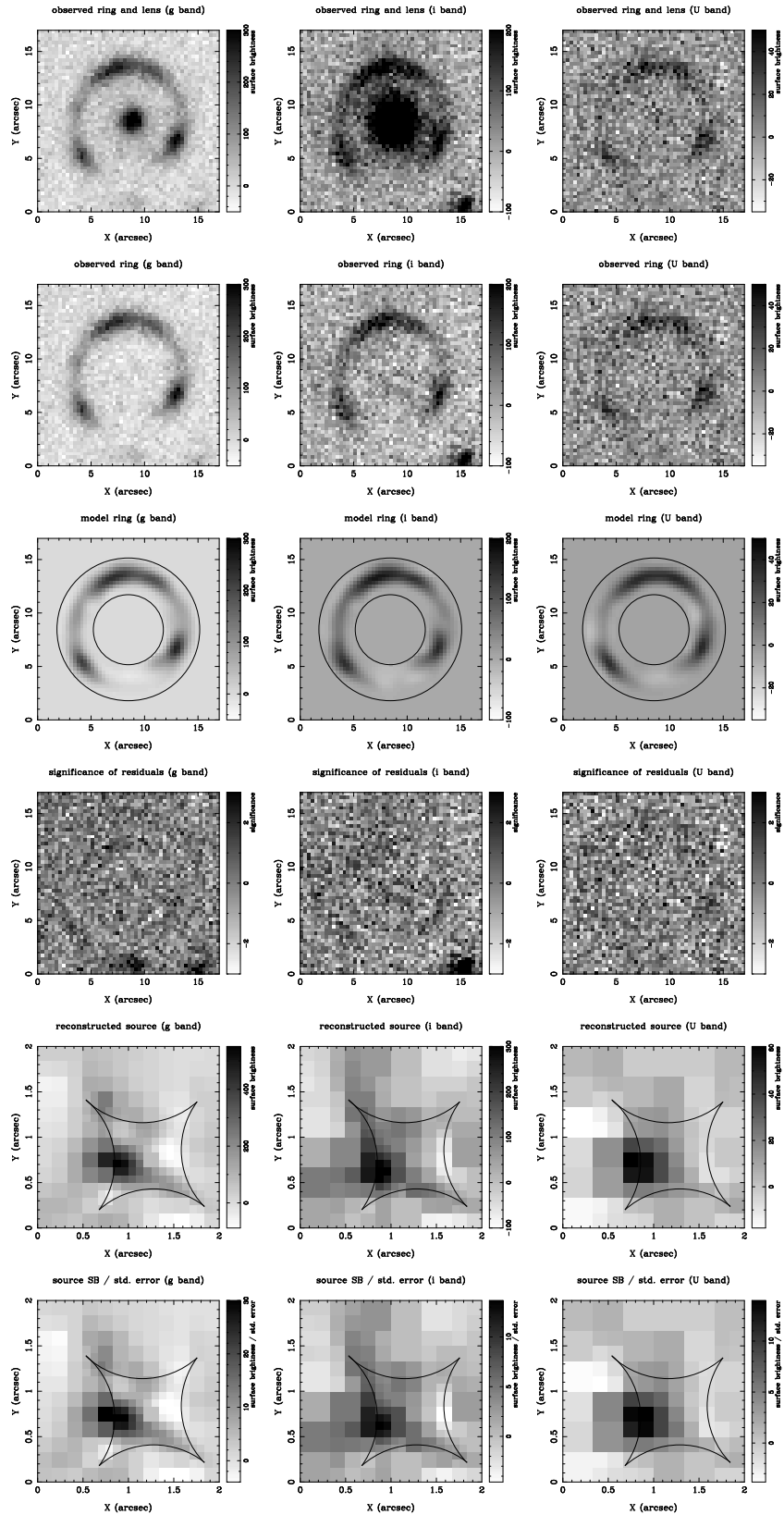}
\hfill}
\epsfverbosetrue
\caption{Image data and source reconstructions. Reading from left to
right, the columns correspond to the $g$, $i$ then $U$ band data.  Top
row gives observed image. Second row shows the lens subtracted
image. Third row is the image of the reconstructed source lensed by
the most probable lens model (the `model image'). The annulus shows the
masked area over which the $\chi^2$ term is evaluated when computing
the evidence.  Fourth row shows the significance of the residuals left
after subtracting the model image from the observed ring image shown
in the second row. Fifth and sixth rows respectively show the
reconstructed source and the source divided by the standard errors
given by the diagonal terms in the covariance matrix $\mathbf{C}$ (see
Section \ref{sec_imp}). The northern source referred to in the text is that
at ($0.7'',1.4''$). Reconstructions for all three bands use the
most probable PL lens model established by the $g$ band data. }
\label{recon}
\end{figure*}

\begin{figure*}
\epsfxsize=16.5cm
{\hfill
\epsfbox{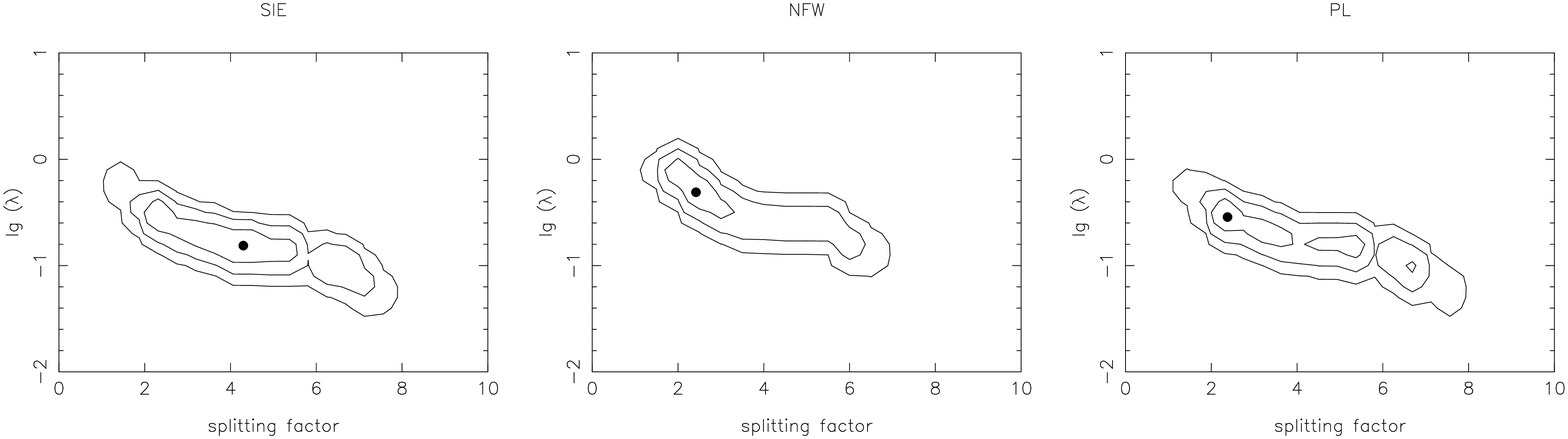}
\hfill}
\epsfverbosetrue
\caption{68\%, 95\% and 99.7\% confidence regions on the splitting factor
and regularisation weight, $\lambda$, for each of the three lens
models using the $g$ band data. The contours are based on the Bayesian
evidence and show the strong degeneracy between both parameters. Each
plot is normalised to the maximum evidence for that lens model,
indicated by the heavy point. The regularisation weight is scaled such
that a value $\lambda=1$ weights the traces of the matrices
$\mathbf{F}$ and $\mathbf{H}$ in equation (\ref{eq_sl}) equally.}
\label{regwt_split_conts}
\end{figure*}

Figure \ref{regwt_split_conts} shows the confidence regions on the
hyperparameters (the regularisation weight and splitting factor). Each
model has its own preferred combination of splitting factor and
regularisation weight, although they are strongly degenerate. Larger
splitting factors prefer smaller regularisation weights because
increasing the splitting factor results in larger source plane pixels
on average which effectively regularises the solution more heavily.

The source reconstructions for the PL model for each of the three
wavebands are shown in Figure \ref{recon}.  For the $i$ and $U$ band
reconstructions, we fixed the lens model parameters at their most
probable values established from the $g$ band data, but varied the
hyperparameters to maximise the evidence. In this way, the lens model
parameters are set by the higher signal-to-noise $g$ band data, but
the $i$ and $U$ band data are allowed to select the splitting factor
and regularisation weight most appropriate to their information
content.

\begin{table}
\centering
\small
\begin{tabular}{cccc}
\hline
Param. & SIE & NFW & PL \\
\hline
$\kappa_0$ & $2.50\pm0.03$ & $0.118\pm0.002$ & $2.30\pm0.03$\\
$\theta$ & $46.5\pm2.7$ & $55.5\pm3.1$ & $49.2\pm3.0$ \\
$q$ & $0.76\pm0.03$ & $0.89\pm0.02$ & $0.78\pm0.03$ \\
$x_c$ & $-0.12^{\asec}\pm0.04^{\asec}$ & 
        $-0.10^{\asec}\pm0.04^{\asec}$ & $-0.11^{\asec}\pm0.04^{\asec}$ \\
$y_c$ & $0.05^{\asec}\pm0.03^{\asec}$ & 
        $0.04^{\asec}\pm0.03^{\asec}$ & $0.02^{\asec}\pm0.03^{\asec}$ \\
$\alpha$ & -- & -- & $1.96\pm0.02$ \\
ln $\epsilon$ & $-4237.7$ & $-4262.7$ & $-4235.4$ \\
\hline
Param. & SIE$+\gamma$ & NFW$+\gamma$ & PL$+\gamma$ \\
\hline
$\kappa_0$ & $2.58\pm0.03$ & $0.116\pm0.002$ & $2.37\pm0.03$\\
$\theta$ & $49.8\pm2.7$ & $47.9\pm3.1$ & $50.8\pm3.1$ \\
$q$ & $0.81\pm0.02$ & $0.86\pm0.02$ & $0.83\pm0.02$ \\
$x_c$ & $-0.10^{\asec}\pm0.04^{\asec}$ & 
       $-0.09^{\asec}\pm0.04^{\asec}$ & $-0.11^{\asec}\pm0.04^{\asec}$ \\
$y_c$ & $0.03^{\asec}\pm0.03^{\asec}$ & 
       $0.04^{\asec}\pm0.03^{\asec}$ & $0.04^{\asec}\pm0.03^{\asec}$ \\
$\alpha$ & -- & -- & $1.95\pm0.02$ \\
$\gamma$ & $0.017\pm0.005$ & $0.011\pm0.006$ & $0.020\pm0.005$ \\
$\theta_\gamma$ & $38.2\pm9.4$ & $46.1\pm12.4$ & $37.7\pm8.6$ \\
ln $\epsilon$ & $-4239.0$ & $-4272.0$ & $-4240.2$\\
\hline
\end{tabular}
\normalsize
\caption{Most probable parameters obtained by maximising the evidence,
$\epsilon$ for each model. Parameters are: total mass normalisation
$\kappa_0$ (in $10^{10}\, {\rm M}_\odot$ kpc$^{-2}$), orientation in degrees
counter-clockwise with respect to the positive y-axis $\theta$, the
axis ratio $q$ (minor axis divided by major axis), lens centroid in
the image plane in arcseconds offset from the observed light centroid
$(x_c,y_c)$ and the slope for the PL model $\alpha$. The top and
bottom halves of the table respectively correspond to the models
without and with external shear of magnitude $\gamma$ and direction
$\theta_\gamma$.}
\label{tab_params}
\end{table}

The velocity dispersion, $\sigma$, implied by the SIE model is
given by
\be
\sigma^2 = \Sigma_{\rm CR} \, r_E \, G
\ee
where $\Sigma_{\rm CR}$ is the critical surface mass density
\citep[see][for example]{Sc92} and $r_E$ is the Einstein radius
which relates to the SIE model parameters via
\be
(r_{\rm E}/1{\rm kpc}) = 2\kappa_0\Sigma_{\rm CR}^{-1}q^{-1/2} \, .
\ee
This gives $r_{\rm E}=28.4$ kpc
corresponding to a velocity dispersion of $496\pm5$ kms$^{-1}$,
which would make the lens one of the the most massive galaxies so far
known! Nonetheless, this is consistent with the result of Gaussian
fitting to absorption lines in the SAO spectrum by
\citet{belokurov07}, which yielded an estimate of
430$\pm$50\,kms$^{-1}$. Although the spectrum is modest, there is
little doubt that the lens is an extreme object -- colour and
luminosity correlate with velocity dispersion and mass, and the lens
is in the brightest and reddest bins for LRGs.  We emphasise that the
modelling both in this paper and in \citet{belokurov07} does not
explicitly include a velocity dispersion constraint. Hence, it is
reassuring that both investigations have come to similar conclusions
as regards the velocity dispersion of the lensing galaxy. Furthermore,
the consistency between the two measurements implies that the stellar
orbits in the LRG are nearly isotropic.

The results listed in Table \ref{tab_params} show that the presence of
external shear is very minor. Furthermore, the evidence ranks all
models incorporating shear with a lower probability than their
non-sheared equivalent models. The sheared models are penalised by
introducing an extra two parameters that do not bring about a
significant improvement in the fit to the data.

At first, this seems surprising, as the lens is located within a group
or loose cluster. With such an enormous image separation
($10^{\asec}$) required, it is natural to expect a significant
contribution from the environment. Even so, there is another telling
indication that the environment plays only a minor role in the
lensing.  It was already established by \citet{kochanek01} that the
ellipticity of an Einstein ring is proportional to the external
shear. The Cosmic Horseshoe ring is very nearly a perfect circle.
This suggests that any perturbation from the cluster is minimal, as
the mismatch between the orientation of the cluster and the lensing
galaxy would generate shear and hence ellipticity in the ring. The
same point is made in \citet{Sa03} -- a narrow spread in images'
galactocentric distances indicates a small or zero external shear and
moderate galaxy ellipticity. We conclude that almost all the
deflection is indeed provided by one very massive galaxy, with the
group environment playing a very minor role.
 
One curiosity is that, for the best-fitting SIE and PL models, the
axis ratio of the baryonic and dark matter in Table \ref{tab_params}
is smaller than the axis ratio of the light distribution in Table
\ref{tab_lens_light}. There seem to be two possible resolutions of
this difficulty. First, there is a well-known degeneracy between
flattening and external shear. In fact, the very minor contribution
from external shear, as in the models in the lower panel of Table
\ref{tab_params}, is already enough to restore the axis ratios to good
agreement. Second, it is quite likely that the ellipticity of the LRG
varies with radius. Although much deeper imaging is required to
confirm this suggestion, there are nonetheless many local examples of
giant ellipticals whose central regions are rather round, but whose
outer parts are much more elongated. A good example is the nominally
E0 galaxy M87, for which the ellipticity rises to 0.4 in the outer
regions \citep{We97}. If a similar situation applies to the Cosmic
Horseshoe lens galaxy, then the photometry of the inner parts may not
be a good guide to the true shape. This may also provide an
explanation as to why the position angle of the major axis of the best
fit light profile is different from the angle preferred by lens
models.

\begin{figure}
\epsfxsize=7cm
{\hfill
\epsfbox{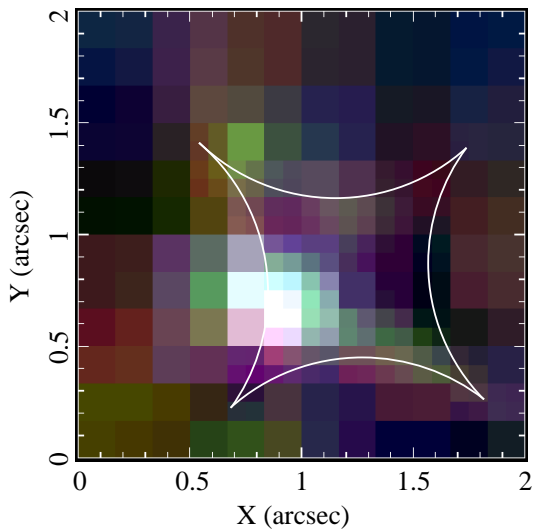}
\hfill}
\epsfverbosetrue
\caption{Colour composite of the reconstructed source. The red, green
and blue channels in this image are formed respectively from the $i$,
$g$ and $U$ band reconstructed sources plotted in the fifth row of
Figure \ref{recon}.}
\label{recon_src_colour}
\end{figure}

Finally, we note from the source reconstructions using the $g$ and $i$
band data in the bottom-most panels of Figure \ref{recon} that there
is evidence for two peaks. However, the secondary northern source does
not appear to be visible in the reconstruction from the noisier $U$
band data. This manifests itself in the colour composite 
source shown in Figure \ref{recon_src_colour}. The red, green and blue
channels of this plot are respectively the $U$, $g$ and $i$ source
surface brightness maps plotted in the fifth row of Figure
\ref{recon}. The northern source has a yellowish-green colour owing to
the lack of $U$ band flux. Unfortunately, it is impossible to say
whether this is an intrinsic colour variation or due to the lower
sensitivity of the $U$ band data. Similarly, the differing source
resolutions between bands prevents a clear interpretation of
the colour of visible structures.

\begin{figure}
\epsfxsize=7cm
{\hfill
\epsfbox{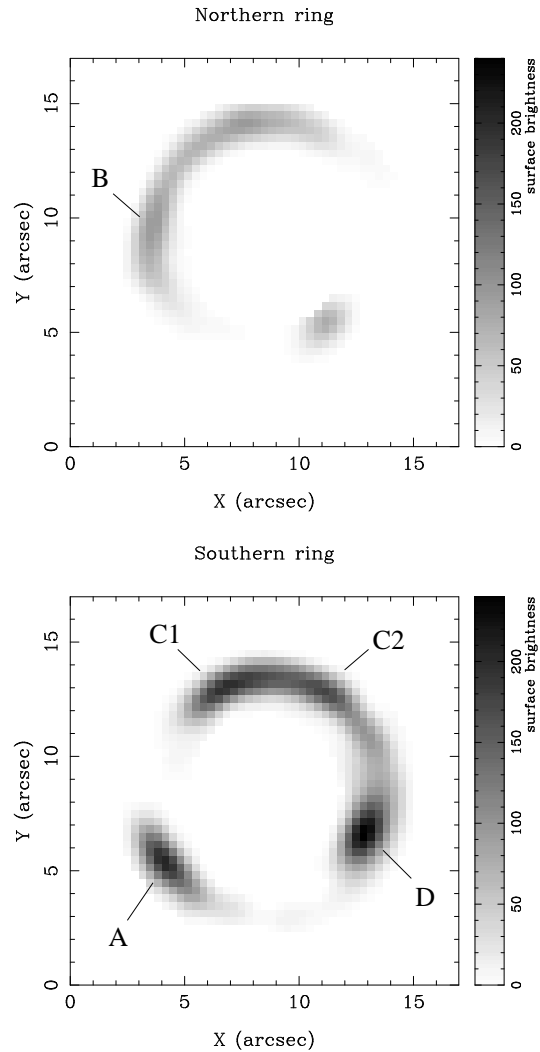}
\hfill}
\epsfverbosetrue
\caption{Images of the northern (top) and southern (bottom)
reconstructed $g$ band source. The northern ring is formed by imaging
all pixels northwards of the line $y=1.2^{\asec}$ in the reconstructed
$g$ band source shown in Figure \ref{recon} and the southern ring from
all pixels to the south of this line.  The ring maxima A to D follow
the labelling of \citet{belokurov07}.}
\label{ring_parts}
\end{figure}

The two peaks in the reconstructed source may be evidence for
substructure or may indicate two sources at different
redshifts. Figure \ref{ring_parts} shows the contributions to the ring
of the Cosmic Horseshoe made by each source. \citep{belokurov07}
already noted that there were five knots or maxima in the flux density
along the ring, which they labelled A, B, C$_1$, $C_2$ and D. The
southern source is mainly responsible for A, C$_1$, C$_2$ and D,
whilst the effect of the northern source is to provide the additional
maximum at B. As the maximum at B is barely discernible in the $U$
band image, it is no surprise that the reconstructed source in $U$
does not show any bimodality.

\section{Summary}
\label{sec_summary}

This paper has provided the first models of the Einstein ring in the
newly discovered Cosmic Horseshoe gravitational lens. The semi-linear
method of \citet{warren03}, in which the source distribution is
pixelised, remains the technique of choice. For a given parametric
model of the lens, the inversion of the source is linear. Here, we
have exploited the refinement of adaptive gridding introduced by
\citet{dye05} and used the Bayesian evidence formulation of
\citet{suyu06} to discriminate between different parametric models on
an equal basis.

The lens in the Cosmic Horseshoe is a luminous red galaxy (LRG) lying
in a group or loose cluster.  Three different mass distributions were
used to model the total luminous and dark matter in the lens --
namely, an isothermal ellipsoid, a Navarro-Frenk-White profile and a
power-law ellipsoid. The effects of the cluster were represented by
external shear. At least as judged by Bayesian evidence, a power-law
ellipsoid without shear provides the best fit. Specifically, the mass
density falls off like $\rho\propto {\tilde r}^{-1.96\pm0.02}$, where
${\tilde r}$ defines similar concentric ellipses with axis ratio $q
\sim 0.8$.

Remarkably, the contribution of the group to the lensing deflection is
minimal, despite the huge image separation ($10^{\asec}$) in
the Cosmic Horseshoe. This result is consistent with the almost
circular nature of the Einstein ring.  However, it means that almost
all the lensing effect is produced by an enormous LRG -- the velocity
dispersion estimated from the modelling is $\sim 500$ kms$^{-1}$. This
mildly exceeds the velocity dispersion of 430$\pm$50\,kms$^{-1}$
already estimated from a low signal-to-noise spectrum by
\citet{belokurov07}. The lens galaxy appears to be the most massive
LRG ever detected. The source reconstructions using the $g$ and the
$i$ band data is double-peaked, although that built from the noisier
$U$ band data is not. Although the nature of the double-peak remains
unclear, this result is consistent with the pattern of density maxima
seen along the ring.

Large separation lenses are now being routinely discovered by searches
through data from the Sloan Digital Sky Survey. These probe a very
different regime to the smaller separation lenses. Tools such as the
ring inversion algorithm employed here can play a substantial role in
understanding the distribution of matter to large radii in very
massive galaxies.

\section*{Acknowledgements}
SD and VB acknowledge financial support from the Science and Technology
Funding Council (STFC).

\label{lastpage}

\end{document}